\documentclass[12pt]{iopart}

\usepackage{iopams}
\usepackage{amssymb}
\usepackage{amsmath}
\usepackage{color}
\usepackage{harvard}
\usepackage{subfigure}
\usepackage{graphicx}
\usepackage{multirow}

\def\R{{\Bbb R}}

\def\h{{\mathbf h}}

\def\u{\ifmmode \mathbf{u} \fi}

\def\x{\mathbf{x}}

\def\bi{\begin{itemize}} \def\ei{\end{itemize}}
\def\be{\begin{eqnarray*}}
\def\ee{\end{eqnarray*}}

\def\0{{\mathbf 0}}
\newcommand{\beq}{\begin{equation}}
\newcommand{\eeq}{\end{equation}}

\def\Npixel{\aleph_{\tiny{\mbox{pixel}}}}

\begin{document}
	
	\title[]{Automatic detection and segmentation of lumbar vertebra from X-ray images for compression fracture evaluation}
	
	\author{Kang~Cheol~Kim\dag,	Hyun~Cheol~Cho\dag, Tae~Jun~Jang\dag, Jong Mun Choi, MD\ddag\footnote[4]{To whom correspondence should be addressed (ilkh0117@deepnoid.com)}, and Jin Keun Seo\dag}
\address{\dag Department of Computational Science and Engineering,
			Yonsei University, Seoul 03722, South Korea}
\address{\ddag DEEPNOID Inc., Seoul, South Korea}

\begin{abstract}
For compression fracture detection and evaluation, an automatic X-ray image segmentation technique that combines deep-learning and level-set methods is proposed.
Automatic segmentation is much more difficult for X-ray images than for CT or MRI images because they contain overlapping shadows of thoracoabdominal structures including lungs, bowel gases, and other bony structures such as ribs. Additional difficulties include unclear object boundaries, the complex shape of the vertebra, inter-patient variability, and variations in image contrast. Accordingly, a structured hierarchical segmentation method is presented that combines the advantages of two deep-learning methods. Pose-driven learning is used to selectively identify the five lumbar vertebra in an accurate and robust manner. With knowledge of the vertebral positions, M-net is employed to segment the individual vertebra. Finally, fine-tuning segmentation is applied by combining the level-set method with the previously obtained segmentation results.
The performance of the proposed method was validated using clinical data, resulting in center position detection error of $25.35\pm10.86$ and a mean Dice similarity metric of $91.60\pm2.22\%$.

\end{abstract}

\maketitle

\section{Introduction}
Compression fracture usually occurs when osteoporosis patient slip down. Severe osteoporosis can cause it without major traumatic event. Patients with compression fractures generally have symptoms such as back pain, but the symptoms are not always clear. Accurate and rapid diagnosis is essential to ensure that suspicious patients do not miss the right time to treat.

There are a variety of modalities to diagnose compression fractures. A plain lumber X-ray is the frontline imaging examination for the diagnosis of spinal fracture and for monitoring the progression of that. X-rays are generally obtained in the first instance because the procedure is fast, inexpensive, and simple. On the other hand,  X-ray have disadvantages of overlapping shadows of other thoracoabdominal 3-dimensional structures, compared with CT or MRI. In clinical terms, accurate segmentation of the lumber vertebra could assist in accurate compression fracture diagnosis and progress estimation.

Automatic segmentation of lumbar vertebra is desirable because manual segmentation is cumbersome and time-consuming. It can help guide the clinician’s assessment and reduce misdiagnosis caused by human error. However, compared with CT and MRI images, automatic segmentation of the lumbar spine from X-ray images is challenging because of the overlapping shadows of complex 3D structures such as the rib cage. It is difficult to segment the five lumbar vertebra selectively without using anatomical and morphological information.

Various automated vertebral segmentation methods have been developed for use with medical imaging modalities, most commonly for CT and to a lesser degree for X-ray images. Most of the methods are based on variants of deformable models \cite{Caselles1988,Cootes1995,Davatzikos2002}, with some constraints to improve accuracy and robustness. Klinder {\it et al.} \cite{Klinder2009} developed a mean-shape-constrained deformable model for CT images, and Ibragimov et al. \cite{Ibragimov2014,Ibragimov2017} developed a landmark-assisted deformable model for CT images that combines the advantages of landmark detection and deformable models with Laplacian shape-editing into a supervised multi-energy segmentation framework. Lim {\it et al.} \cite{Lim2013} integrated an edge-mounted Willmore flow with prior shape energies into a level-set framework for the segmentation of spinal vertebra from CT images.
Kadoury {\it et al.} \cite{Kadoury2011,Kadoury2013} used an articulated deformable model for spine segmentation in CT, where manifold embeddings are used to infer constellations accounting for deformations, and higher-order Markov random fields are used to infer articulated objects directly from low-dimensional parameters. Glocker {\it et al.} \cite{Glocker2012,Glocker2013} developed a method based on regression forests for the rough detection of vertebra and probabilistic graphical models for accurate localization and identification of individual vertebra in CT. For a comprehensive comparative study of vertebral segmentation in CT, see \cite{Yao2016} and references therein.
With regard to segmentation from X-ray images, Arif {\it et al.} \cite{AlArif2018} developed a deep-learning-based fully automatic framework for segmentation of cervical vertebra.

Difficulties in directly applying existing segmentation methods to lumbar spine X-rays include the multiple overlapping shadows of the ribs and pelvis, relatively weak contrast, and the need to identify the five lumbar vertebra individually. Accordingly, it is necessary to use the local and global characteristics of lumbosacral spine X-rays and consider factors such as the position of the sacrum and the curve of the vertebral column.

This paper proposes fully automatic lumbar vertebral segmentation from X-ray images by combining deep-learning techniques and level-set methods. The proposed method comprises four main steps, as follows.
\begin{enumerate}
   \item Pre-processing of X-ray images by an adoptive histogram equalization, which are used for adoptive contrast enhancement.
   \item Pose-driven learning to identify each of the five lumbar vertebra.
   \item After their positions are known, M-net based segmentation of the individual lumbar vertebra.
   \item Subsequent fine-tuning of the segmentation by combining these deep-learning segmentation results (in the previous steps) with the level-set method.
\end{enumerate}

The performance of the proposed method was validated on clinical X-ray images from 80 normal person and 80 abnormal person. The experimental results show that the proposed method provide reasonable performance for localization and segmentation of lumbar vertebra. We achieved the center position detection error of $25.35\pm10.86$, and $91.60\pm2.22\%$ Dice similarity metric for segmentation of the five lumbar vertebra.

\section{Methods}
This section proposes a fully automated method for segmentation of the five lumbar vertebra from X-ray images. Segmentation from X-ray images is complicated by the overlapping shadows of other thoracoabdominal 3-dimensional structures.
In addition, lateral view X-ray images usually include the thoracic spines which are adjacent to each other and have similar shape.

\begin{figure}[!h]
	\centering
	{\includegraphics[width=0.95\textwidth]{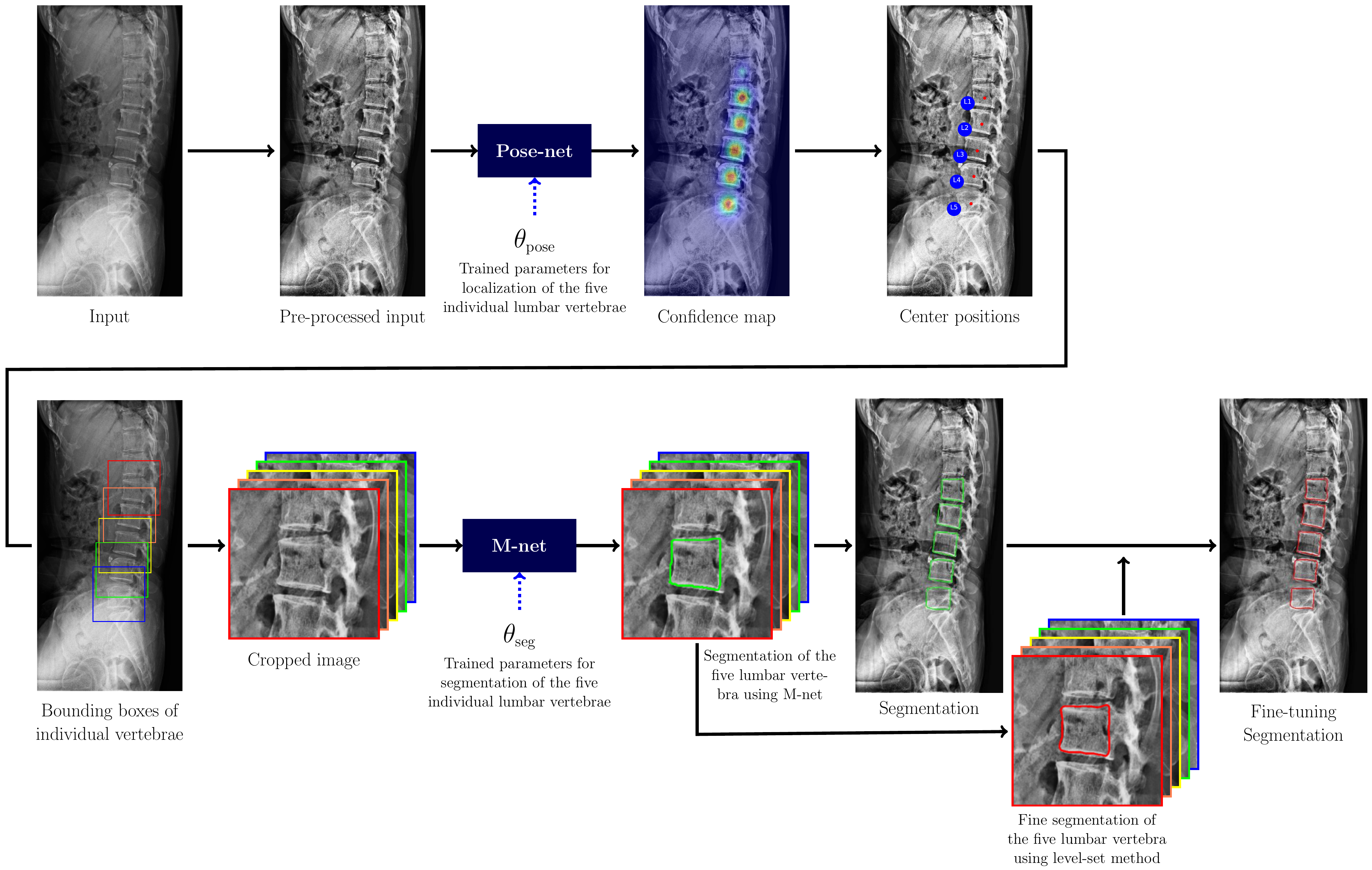}}
	\caption{An overview of the proposed method. In the training phase, we train the two different neural networks to train parameters $\theta_{\mbox{\tiny{pose}}}$ and $\theta_{\mbox{\tiny{seg}}}$, respectively: (1) a pose-driven deep learning network for identification of each lumbar vertebra, and (2) a segmentation network for fine segmentation of individual vertebra. The trained parameters are used in the testing phase. \label{fig:overall_process}}
\end{figure}

To address these problems, a three-part hierarchical method is adopted that mimics the steps in the clinician's decision process: spine localization, segmentation of lumbar vertebra, and fine-tuning of segmentation. The overall process is shown in Fig. \ref{fig:overall_process}. A landmark detection method is used to identify the center of the lumbar spine. From the knowledge of the central position, we extract bounding boxes corresponding to the five individual lumbar vertebra. Deep-learning-based segmentation was then applied to identify the vertebral levels of the extracted patches, and the level-set method was subsequently used to improve the quality of segmentation.

\subsection{Pre-processing}

\begin{figure}[!h]
	\centering
	{\includegraphics[width=0.85\textwidth]{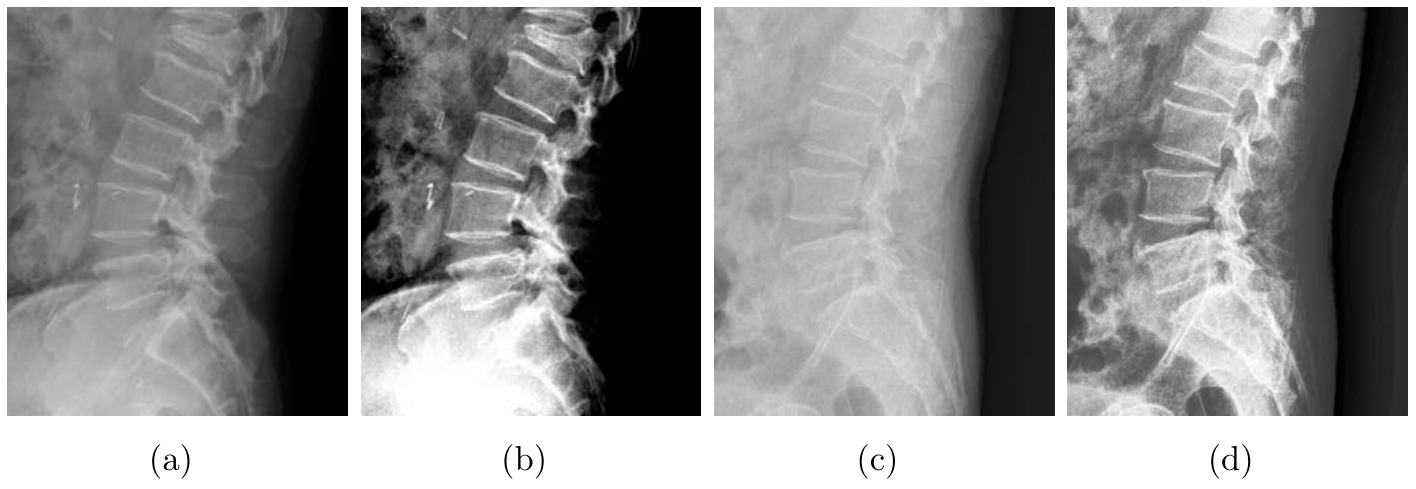}}
	\caption{Pre-processing X-ray images. Original X-ray images are shown in (a) and (c). We used the histogram equalization method to (a) and (c) in order to increase the contrast of images, and results are shown in (b) and (d), repectively.}
	\label{fig:pre-processing}
\end{figure}

Given that X-ray images have a narrow intensity distribution, an adaptive histogram equalization method \cite{Pizer1987}  is employed in pre-processing to increase the contrast of the images by spreading the intensity values. Then, Gaussian filtering is applied to the contrast-enhanced images to alleviate the background noise.
Fig. \ref{fig:pre-processing} for these preprocessing steps images.

Throughout this paper, $I_o(\x)\in \mathbb{R}^{3072\times1536}$ represents an image obtained by applying the preprocessing procedure, where $\x=(x_1,x_2)$ denotes the pixel position. Each image $I_o$ is resized to $512\times256$ pixels, and the resized image, denoted by $I(\x)$,  is used as input of the pose-estimation method described in the next section.

\subsection{Localization of the five lumbar vertebra}

This section describes how to automatically find the center positions of five lumber vertebra (lower back)  between rib cage and the pelvis, which are denoted L1 to L5, starting at the top. Selective detection of the five L-vertebra is a difficult task because of the need to observe neighboring structures such as the sacrum and the whole spine. To solve this problem, we employed a deep learning-based pose estimation method \cite{Toshev2014,Wei2016,Cao2017}, which is widely used in computer vision area for detecting human joint.

\begin{figure}[!h]
	\centering
	{\includegraphics[width=0.95\textwidth]{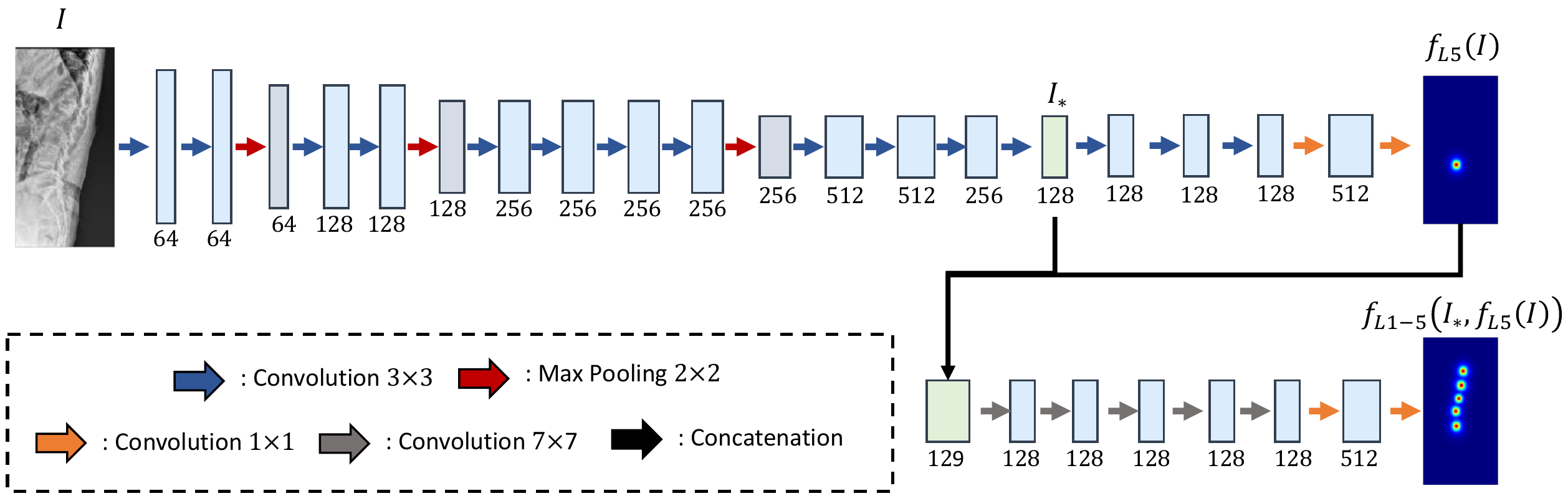}}
    \caption{Architecture of the proposed Pose-net for localizing the five lumbar vertebra. The function $f_{\mbox{\tiny{L5}}}$ returns the confidence map providing the belief of the center of L5 vertebra. The output of $f_{\mbox{\tiny{L1-5}}}$ is the confidence map providing the belief of the all centers of five lumbar vertebra.}
	\label{fig:pose_net}
\end{figure}

The pose estimation aims to predict the pose of five lumber vertebra in the image $I$, where the output poses are expressed by a vector ${\bf P}=({\bf p}_1,\cdots, {\bf p}_5)\in \R^{2\times5}$, representing the set of center positions of five lumbar vertebra. To achieve this, we use two-stage neural networks, namely Pose-net denoted by functions $f_{{\mbox{\tiny{L5}}}}$ and $f_{{\mbox{\tiny{L1-5}}}}$, to generate two confidence maps; the first confidence map $y_{{\mbox{\tiny{L5}}}}=f_{{\mbox{\tiny{L5}}}}(I)\in \mathbb{R}^{512\times 256}$ provides the belief of the center of L5 vertebra and the second confidence map $y_{\mbox{\tiny{L1-5}}}=f_{\mbox{\tiny{L1-5}}}(I_{*},y_{{\mbox{\tiny{L5}}}})\in \mathbb{R}^{512\times 256}$ provides the belief of ${\bf P}$ by taking advantage of the first one $y_{{\mbox{\tiny{L5}}}}$.
Here, $I_*$ is an intermediate feature layer of $f_{\mbox{\tiny{L5}}}$ with input $I$ as shown in Fig. \ref{fig:pose_net}.
We adopt a convolutional neural network(CNN) to learn functions $f_{\mbox{\tiny{L5}}}$ and $f_{\mbox{\tiny{L1-5}}}$.
The input of $f_{\mbox{\tiny{L5}}}$ is $I$ and $f_{\mbox{\tiny{L5}}}(I)$ is expressed as
\begin{equation}\label{eq:pose_function1}
  f_{\mbox{\tiny{L5}}}(I)=W^{l}\circledast \eta(\cdots\mathcal{P}(\eta(W^{2}\circledast(\eta(W^{1}\circledast I)))))
\end{equation}
where $W\circledast \h$ is the convolution of $\h$ with the weight $W$; $\mathcal{P}$ is the max pooling; and $\eta$ is the rectified linear unit activation function $ReLU$.
The input of $f_{\mbox{\tiny{L1-5}}}$ is a concatenated vector of $y_{\mbox{\tiny{L5}}}$ and $I_{*}$. Fig. \ref{fig:pose_net} shows the architecture of $f_{\mbox{\tiny{L1-5}}}$.

These networks $f_{\mbox{\tiny{L5}}}$ and $f_{\mbox{\tiny{L1-5}}}$ are learned simultaneously, using the training data $\mathcal{S}_{\mbox{\tiny pose}}:=\{ I^{(n)},y_{\mbox{\tiny{L5}}}^{(n)}, y_{\mbox{\tiny{L1-5}}}^{(n)}\}_{n=1}^N$.
The loss function is given by
\begin{equation}\label{eq:loss1_pose1}
	\mathcal{L}(\theta_{\mbox{\tiny{pose}}}) = \frac{1}{N} \sum_{n=1}^{N} \mathcal{L}^{(n)}(\theta_{\mbox{\tiny{pose}}})
\end{equation}
where $\theta_{\mbox{\tiny{pose}}}$ is a set of all parameters in the network and $\mathcal{L}^{(n)}(\theta_{\mbox{\tiny{pose}}})$ is the sum of the intermediate loss and the final loss:
\begin{equation}\label{eq:loss1_pose2}
	\mathcal{L}^{(n)}(\theta_{\mbox{\tiny{pose}}}) =
	\left \| f_{\mbox{\tiny{L5}}}(I^{(n)}) - y_{\mbox{\tiny{L5}}}^{(n)} \right \|^2_2 + \left \| f_{\mbox{\tiny{L1-5}}}({I_{*}}^{(n)},f_{\mbox{\tiny{L5}}}(I^{(n)})) - y_{\mbox{\tiny{L1-5}}}^{(n)} \right \|^2_2.
\end{equation}

Here, the labeled data $y^{(n)}_{\mbox{\tiny{L5}}}$ is given by
\begin{equation}\label{eq:confidence_gt1}
	y^{(n)}_{\mbox{\tiny{L5}}}(\mathbf{\x})=\exp{\left( -\frac{||\mathbf{\x}-{\bf p}_5||_2^2}{\sigma^2} \right)}
\end{equation}
where ${\bf p}_5$ is the ground-truth of the center position of L5 vertebra and $\sigma^2$ is given by 1/4 of L5 vertebra height in the image $I^{(n)}$. The others ($y_{\mbox{\tiny{L1}}}, \cdots, y_{\mbox{\tiny{L4}}}$) are given in the same way. Then the confidence map $y_{\mbox{\tiny{L1-5}}}$ can be obtained by
\begin{equation}\label{eq:confidence_gt2}
	y_{\mbox{\tiny{L1-5}}}(\x)=\max\{y_{\mbox{\tiny{L1}}}(\x), \cdots, y_{\mbox{\tiny{L5}}}(\x)\}.
\end{equation}

\begin{figure}[!h]
	\centering
    {\includegraphics[width=0.85\textwidth]{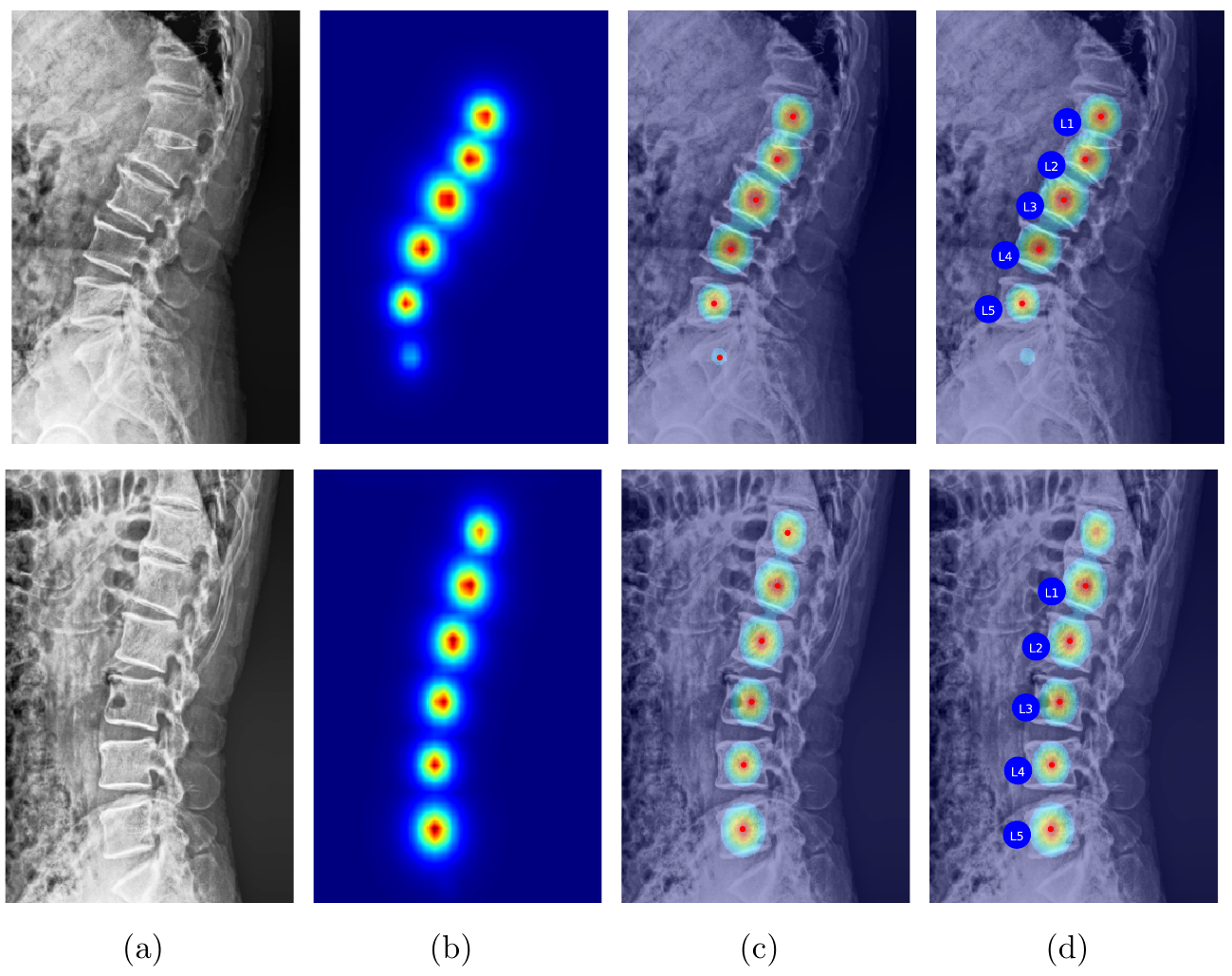}}
	\caption{Localization of the five lumbar vertebra. (a) Test image $I$. (b) The confidence map $y_{\mbox{\tiny{L1-5}}}(\x)=f_{\mbox{\tiny{L1-5}}}({I_{*}},f_{\mbox{\tiny{L5}}}(I))$. (c) Local maxima denoted by red dots. (d) Determination of the center positions of the five lumbar vertebra. In the top row, the wrong candidates (whose score are less than half of the mean score) are removed. In the bottom row, the wrong candidate (i.e., the red dot at the top in (c)) is still alive even with the mean score filtering. The five candidates starting from the bottom candidate are selected as shown in (d).}
	\label{fig:NMS}
\end{figure}

Note that the functions $f_{\mbox{\tiny{L5}}}$ and $f_{\mbox{\tiny{L1-5}}}$ are determined by minimizing the loss function in (\ref{eq:loss1_pose1}) using the training data. Hence, given a test data $I$, this Pose-net provides the confidence map $y_{\mbox{\tiny{L1-5}}}(\x)=f_{\mbox{\tiny{L1-5}}}({I_{*}},f_{\mbox{\tiny{L5}}}(I))$, as shown in Fig. \ref{fig:NMS} (a).

Now, we are ready to explain our method to determine the center positions ${\bf P}=({\bf p}_1,\cdots, {\bf p}_5)\in \R^{2\times5}$ of five lumbar vertebra from this confidence map $y_{\mbox{\tiny{L1-5}}}(\x)$.
We first applied the Otsu's thresholding \cite{Otsu1979} to the confidence map $y_{\mbox{\tiny{L1-5}}}(\x)$ in Fig. \ref{fig:NMS}(a) to remove small local perturbations in $y_{\mbox{\tiny{L1-5}}}(\x)$ so that the local maxima distant from the spines are filtered. See Fig. \ref{fig:NMS} (c), which shows six local maxima in the image. These local maximum points are the candidates of the center positions.
Next, we need to select five points ${\bf P}=({\bf p}_1,\cdots, {\bf p}_5)\in \R^{2\times5}$ from the several candidates.
To do this, we computed the score by averaging the pixel values of the image $y_{\mbox{\tiny{L1-5}}}(\x)$ in a window of size $31\times31$, centered at each local maximum point. We excluded the candidates whose score are less than half of the mean score.
Finally, we select the five candidates starting from the bottom candidate, as shown in Fig. \ref{fig:NMS} (d).

\subsection{Deep learning-based segmentation of lumbar vertebra}
Our segmentation method takes advantage of the knowledge of the center positions ${\bf P}$  of the five vertebra (that are obtained from the Pose-net explained in the previous section) to greatly reduce the area performing the segmentation.
The segmentation is performed in the original high-resolution image $I_o\in \mathbb{R}^{3072\times1536}$ instead of the resized low-resolution image $I\in\mathbb{R}^{512\times256}$. Given ${\bf P}$ in the resized image $I$, it is easy to calculate the corresponding vector ${\bf P}_o=({\bf p}_{o,1},\cdots, {\bf p}_{o,5})\in \R^{2\times5}$ representing the center positions of the five vertebra in the original image $I_o$.

\begin{figure}[!h]
	\centering
	{\includegraphics[width=0.9\textwidth]{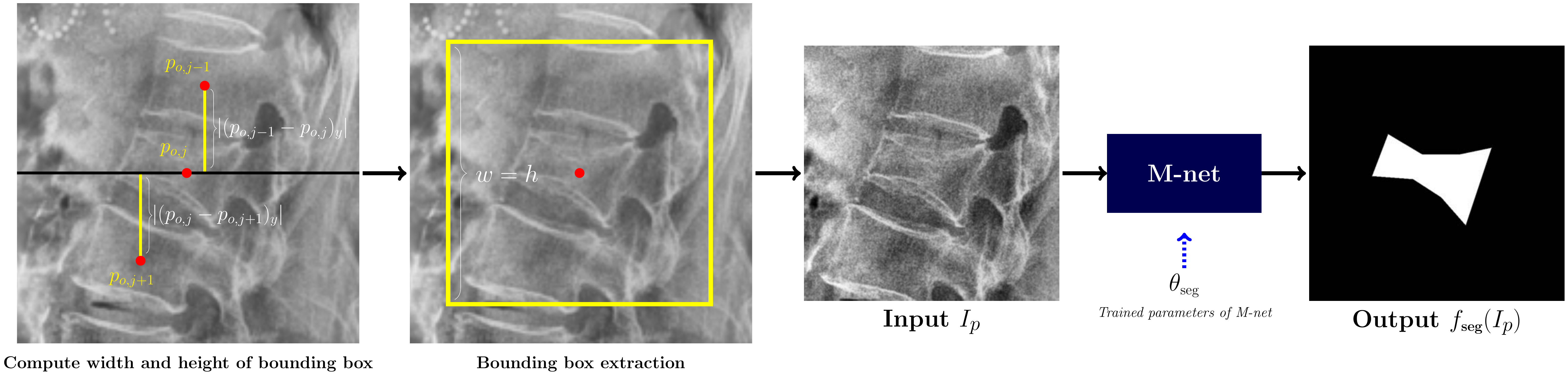}}
    \caption{Segmentation process. Our segmentation method utilizes the five spine positions  to greatly reduce the area (the contents of the bounding boxes) of performing the segmentation. Then, M-net was applied to segment the five lumbar vertebra.}
	\label{fig:segmentation_process}
\end{figure}

The proposed segmentation method takes as input the content of a bounding box of each lumbar vertebra, as shown in Fig. \ref{fig:segmentation_process}, where the five bounding boxes are centered at ${\bf P}_o$, and the height $h$ and width $w$ of which are the same:
\begin{equation}\label{eq:bbox_wh}
	h=w=
	\begin{cases}
	3|({\bf p}_{o,j}-{\bf p}_{o,j+1})_y|, & \text{for } j=1\\
	\frac{3}{2}\left(|({\bf p}_{o,j-1}-{\bf p}_{o,j})_y|+|({\bf p}_{o,j}-{\bf p}_{o,j+1})_y|\right), & \text{for } j=2,3,4\\
	3|({\bf p}_{o,j-1}-{\bf p}_{o,j})_y|, & \text{for } j=5
	\end{cases}
\end{equation}
where the subscript $y$ stands for the vertical component of the corresponding vector.
See Fig. \ref{fig:segmentation_process} for a description of the bounding box.

We use the M-net \cite{Fu2018} to learn the segmentation map $f_{\mbox{\tiny seg}} :{I_{p}} \mapsto y_{\mbox{\tiny seg}}$, where $I_p$ denotes the content of a bounding box ($I_p$ is a resized image to $224\times224$) and $y_{\mbox{\tiny seg}}$ is a binary image representing vertebra segmentation corresponding to the patch $I_p$.
The M-net is based on U-net \cite{Ronneberger2015} and has advantages by adding two major parts: (1) multi-scale layer used to construct an image pyramid input and (2) multi-label loss function with side-output layer to learn local and global information at the same time. Here, the multi-scale input is to integrate multiple level receptive field\cite{Fu2018}, and the multi-label loss in (\ref{eq:loss1_seg}) can deal with the vanishing gradient problem by replenishing back-propagated gradients \cite{Wei2016}.

\begin{figure}[!h]
	\centering
	{\includegraphics[width=0.9\textwidth]{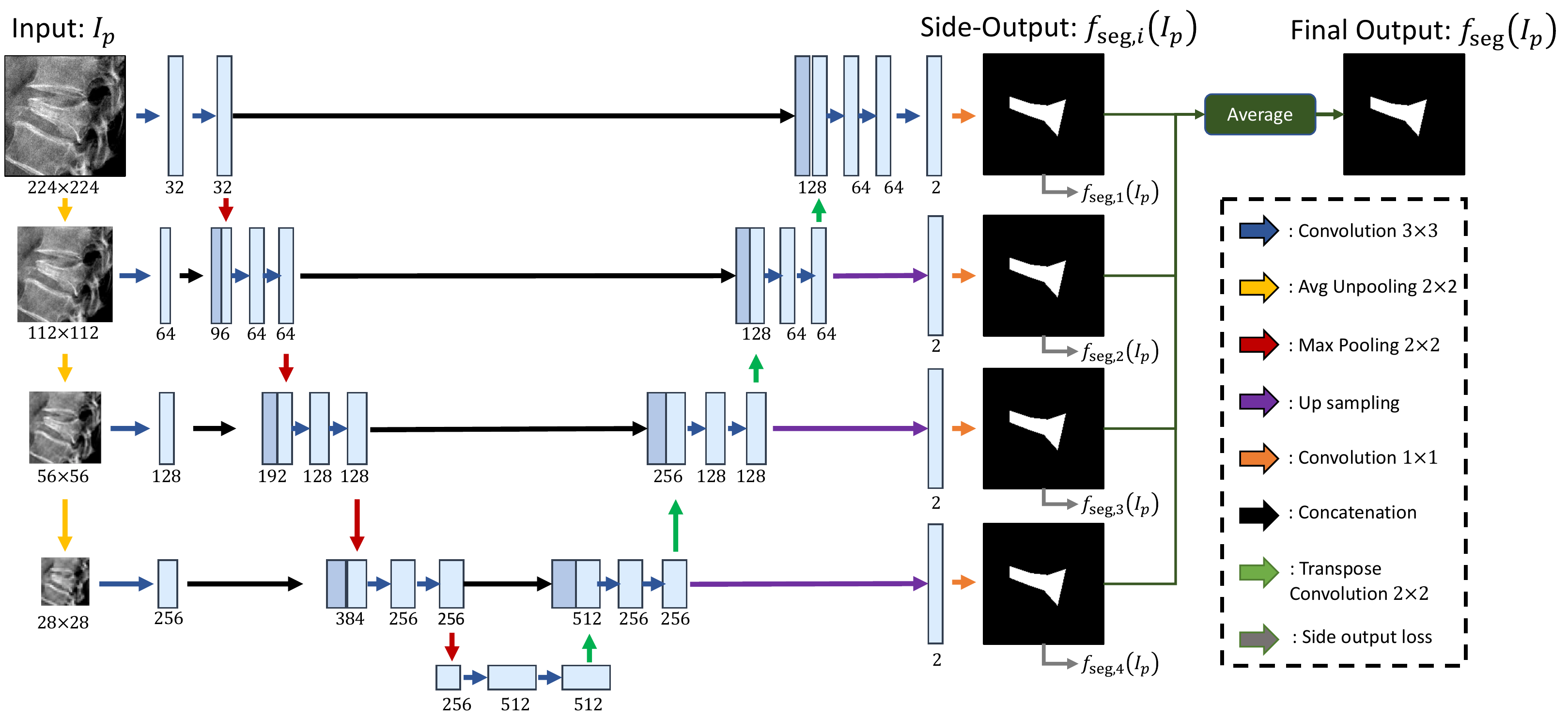}}
    \caption{M-net architecture to learn local and global structure at the same time. The multi-scale input layer constructs an image pyramid to achieve the multiple level receptive field\cite{Fu2018}.}
    \label{fig:segmentation_network}
\end{figure}

Fig. \ref{fig:segmentation_network} shows the M-net structure, and $f_{\mbox{\tiny seg}}$ is expressed as
\begin{equation}
	f_{\mbox{\tiny seg}}(I_p;\theta_{\mbox{\tiny{seg}}})=
\frac{1}{4}\sum_{i=1}^{4}f_{\mbox{\tiny seg},i}(I_p;\theta_{\mbox{\tiny{seg}},i})
\end{equation}
where $f_{\mbox{\tiny seg},i}$ is the function producing the $i$-th side output. Here, $\theta_{\mbox{\tiny{seg}}} = (\theta_{\mbox{\tiny{seg}},1},\cdots,\theta_{\mbox{\tiny{seg}},4})$ is a set of parameter related to $f_{\mbox{\tiny seg},i}$ for $i=1,\cdots,4$.
The M-net $f_{\mbox{\tiny{seg}}}$ is learned using training data $\mathcal{S}_{\mbox{\tiny seg}}:=\{ I_p^{(n)},y_{\mbox{\tiny{seg}}}^{(n)}\}_{n=1}^N$. The loss multi-label function is given by

\begin{equation}\label{eq:loss1_seg}
  \mathcal{L}(\theta_{\mbox{\tiny{seg}}})=\frac{1}{4N} \sum_{n=1}^{N} \sum_{i=1}^{4} \left[ -\frac{1}{\Npixel} \left< y_{\mbox{\tiny seg}}^{(n)}, {\mathbb L}\left ({f_{\mbox{\tiny seg},i}(I^{(n)}_p;\theta_{\mbox{\tiny{seg}},i})}\right)\right> \right]
\end{equation}
where $\Npixel$ denotes the number of pixels of the input image, $<\cdot,\cdot>$ denotes an inner product, and ${\mathbb L}(\cdot)$ denotes an element-wise log operation.
The segmentation function $f_{\mbox{\tiny seg}}$ is obtained by minimizing loss in (\ref{eq:loss1_seg}).

\subsection{Fine-tuning of segmentation}
For a fine-tuning of segmentation, one may use the level-set method \cite{Kass1987,Caselles1993,Malladi1995,Sussman1994,Osher2001} with using the previously obtained segmentation results. Given an X-ray image patch $I_p$ and deep learning-based segmentation $y_{\mbox{\tiny seg}}$, the following energy functional is used to provide a fine segmentation:
\begin{equation}
	\begin{split}\label{eq:level_set_energy}
		\Phi(\varphi)&=\int_{\Omega} g(\x) \delta(\varphi(\x)) | \nabla \varphi(\x) | d\x + \frac{1}{2}\int_{\Omega} \left( | \nabla \varphi(\x) |-1 \right)^2 d\x \\
	&+ \int_{\Omega} g(\x)H(-\varphi(\x))d\x +\lambda \int_{\Omega} y_{\mbox{\tiny seg}}(\x)H(-\varphi(\x))d\x
	\end{split}
\end{equation}
where $\varphi$ is a level set function, $H$ is the one-dimensional Heaviside function, $G_\sigma$ is Gaussian kernel, $g(\x)=\frac{1}{1+|\nabla G_\sigma \ast I_p(\x)|^2}$ is an edge detector, $\delta$ is a regularized delta function, and $y_{\mbox{\tiny seg}}$ is the binary image obtained by M-net in the previous section. A segmented region $\{ \x: \varphi (\x) <0\}$ is obtained via a level set function $\varphi$ which minimize the energy functional in (\ref{eq:level_set_energy}).
To compute a minimizer $\varphi$ for the energy functional $\Phi(\varphi)$, the following parabolic equation is solved to get a static state:
\begin{equation}
	\begin{split}\label{eq:level_set_derivative}
		\frac{\partial}{\partial t} \varphi(\x)&= \delta(\varphi)\left[ \left( \nabla g\cdot\frac{\nabla \varphi}{| \nabla \varphi |}+\nabla \cdot \frac{\nabla \varphi}{| \nabla \varphi |} g \right) | \nabla \varphi | \right]  \\
		&+ \left[ \nabla^2 \varphi - \nabla \cdot \frac{\nabla \varphi}{| \nabla \varphi |} \right] + g \delta(\varphi) +\lambda y_{\mbox{\tiny seg}} \delta(\varphi).
	\end{split}
\end{equation}

\begin{figure}[!h]
	\centering
	\subfigure[]{\includegraphics[width=0.4\textwidth]{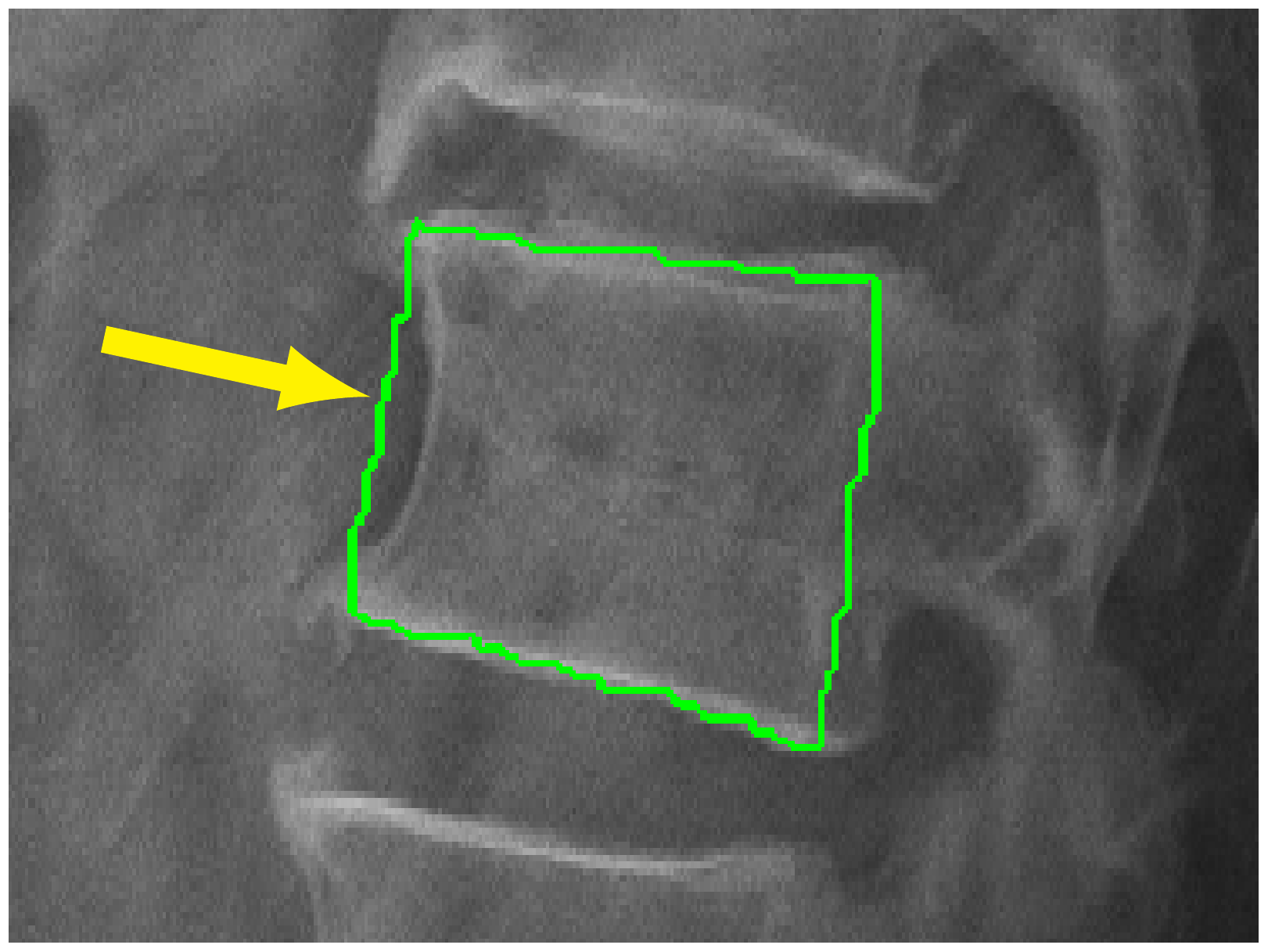}}
    \subfigure[]{\includegraphics[width=0.4\textwidth]{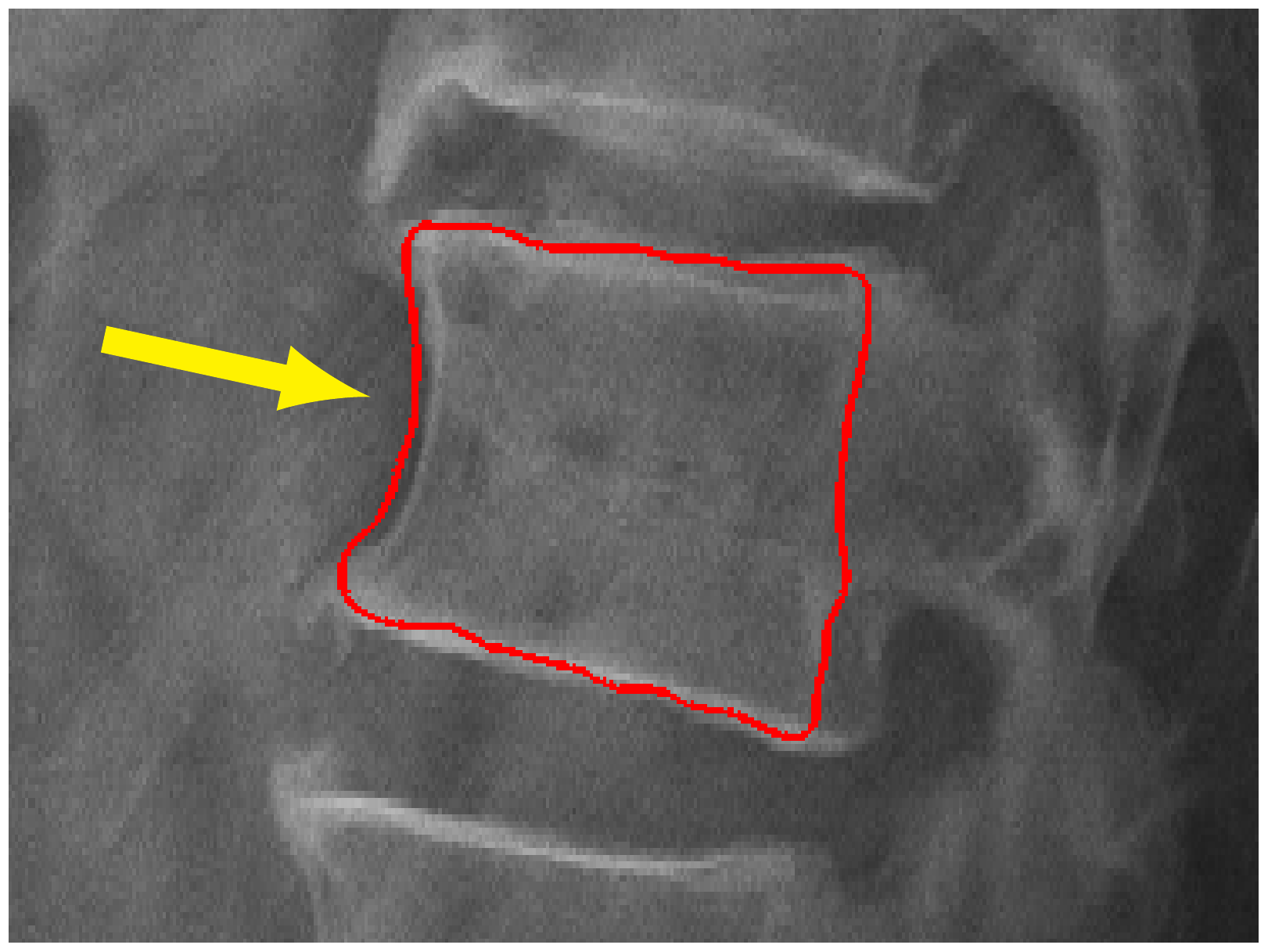}}
	\caption{Fine-tuning segmentation method. (a) Segmentation by M-net. (b) Fine-tuning segmentation using the level-set approach in (\ref{eq:level_set_energy}).}
	\label{fig:level_set_method}
\end{figure}

In the last term of \eqref{eq:level_set_energy}, Chan-Vese method \cite{Chan2001} is applied to the binary image $y_{\mbox{\tiny seg}}$, which was used as the initial segmentation as well as a strong fidelity to a target segmentation.
The key role of the last term is that the level set of a minimizer $\phi$ of \eqref{eq:level_set_energy} is very close to the edge of $y_{\mbox{\tiny seg}}$. For the first three terms of \eqref{eq:level_set_energy} \cite{Li2010}, a distance regularization term and an external energy are used to push the contour to a target area (\ref{eq:level_set_energy}). See Fig. \ref{fig:level_set_method} for the fine-tuning of segmentation.
One may use other level set methods for fine-tuning segmentation.

\section{Experiments and Results}

In this experiments, Python with Tensorflow was used to implement deep learning framework and MATLAB was used for data processing. All process was performed in workstation equipped with the two Intel(R) Xeon(R) E5-2630 v4 @ 2.20GHz CPU, 128GB DDR4 RAM, and 4 NVIDIA GeForce GTX 1080ti 11GB GPU.
\subsection{Data}
\begin{figure}[!h]
	\centering
	{\includegraphics[width=0.9\textwidth]{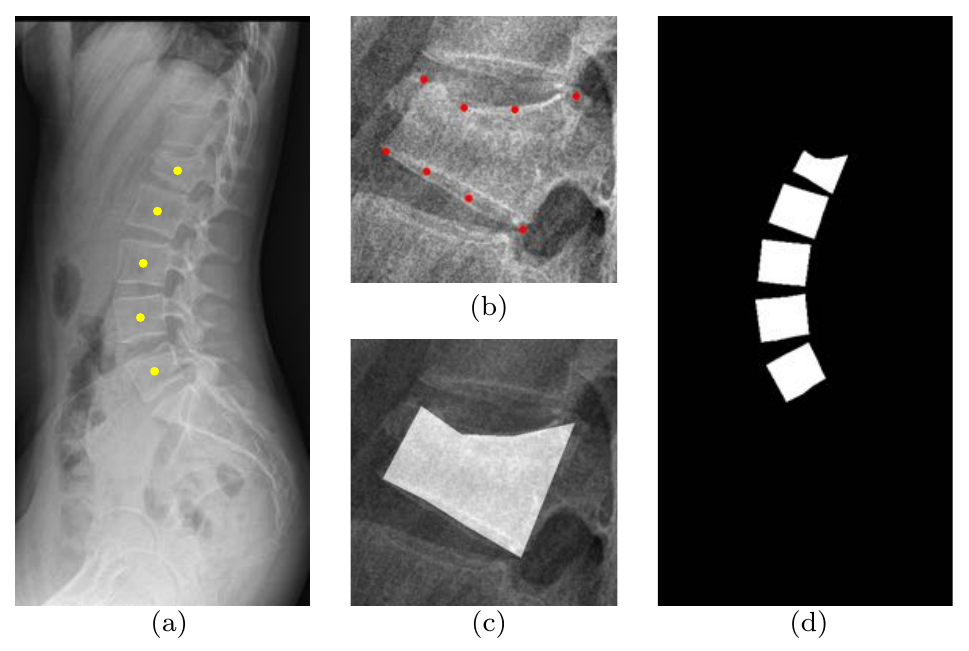}}
    \caption{Labeled training data. Given X-ray image, we manually labeled the center positions (a). Segmentation label of the lumbar vertebra (c) was given by plotting 8 points, as shown in (b). Segmentation label of the five lumbar vertebra was shown in (d).}
	\label{fig:data_processing}
\end{figure}

The training data are 637 Digital Radiography(DR) X-ray images, and test data are 160 X-ray images which consist of DR and Computed Radiography(CR) X-ray images. In training process, we split the training data into 537 and 100 for training and validation, respectively. The size of X-ray images were approximately $3000\times1500$ and we resized the images to $3072\times1536$.
We first manually labeled the center positions denoted by yellow points in Fig. \ref{fig:data_processing} (a).
Segmentation label of the lumbar vertebra(Fig. \ref{fig:data_processing} (c)) was given by plotting 8 red points(Fig. \ref{fig:data_processing} (b)).
The segmentation label of the five lumbar vertebra was shown in Fig. \ref{fig:data_processing} (d).

Data processing and augmentation method of training data are explained as following:
\begin{enumerate}
	\item For the training data of Pose-net, we resized the all images to $512\times256$. Then using (\ref{eq:confidence_gt1}) and (\ref{eq:confidence_gt2}), we computed ground-truth confidence map using center ground-truth positions $P$.
	\item For the training data of M-net, we first extracted the patches of input image and segmentation label using center positions ${\bf P}_o$. Then we resized all extracted patches to $224\times224$.
	\item In the patch extraction process, random noise was added to ${\bf P}_o$ to reflect the errors of center positions which occur during the test stage.
	\item For the augmentation of data, we applied the random contrast adjustment, random cropping, and random rotation within angle $-15^{\circ}$ to $15^{\circ}$.
\end{enumerate}

\subsection{Training and validation of the proposed networks}
The training of the proposed networks are carried out by minimizing loss functions in (\ref{eq:loss1_pose1}) and (\ref{eq:loss1_seg}) using Adam method \cite{Kingma2014}. Here, the batch size was selected to 4 in the consideration of our computational ability. We used the batch normalization \cite{Ioffe2015} which allows higher learning rate, resulting in relatively short training time. We set the learning rate to $10^{-3}$. For initial 50 epoch, we used warm-up learning rate \cite{He2015,Baumgartner2017} of $10^{-5}$ to prevent rapid decrease of training loss.
\begin{figure}[!h]
	\centering
	\subfigure[]{\includegraphics[height=0.32\textwidth]{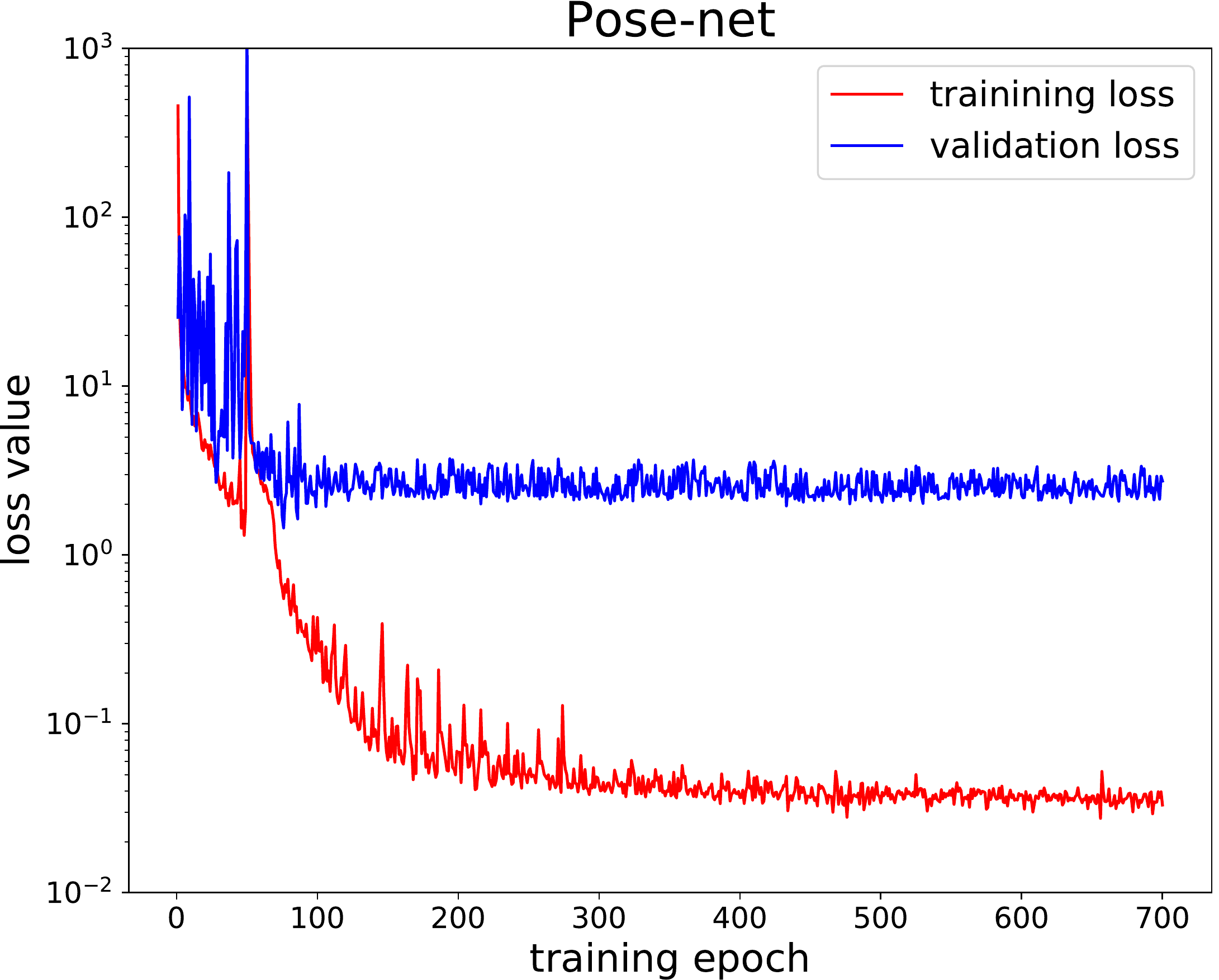}}
    \subfigure[]{\includegraphics[height=0.32\textwidth]{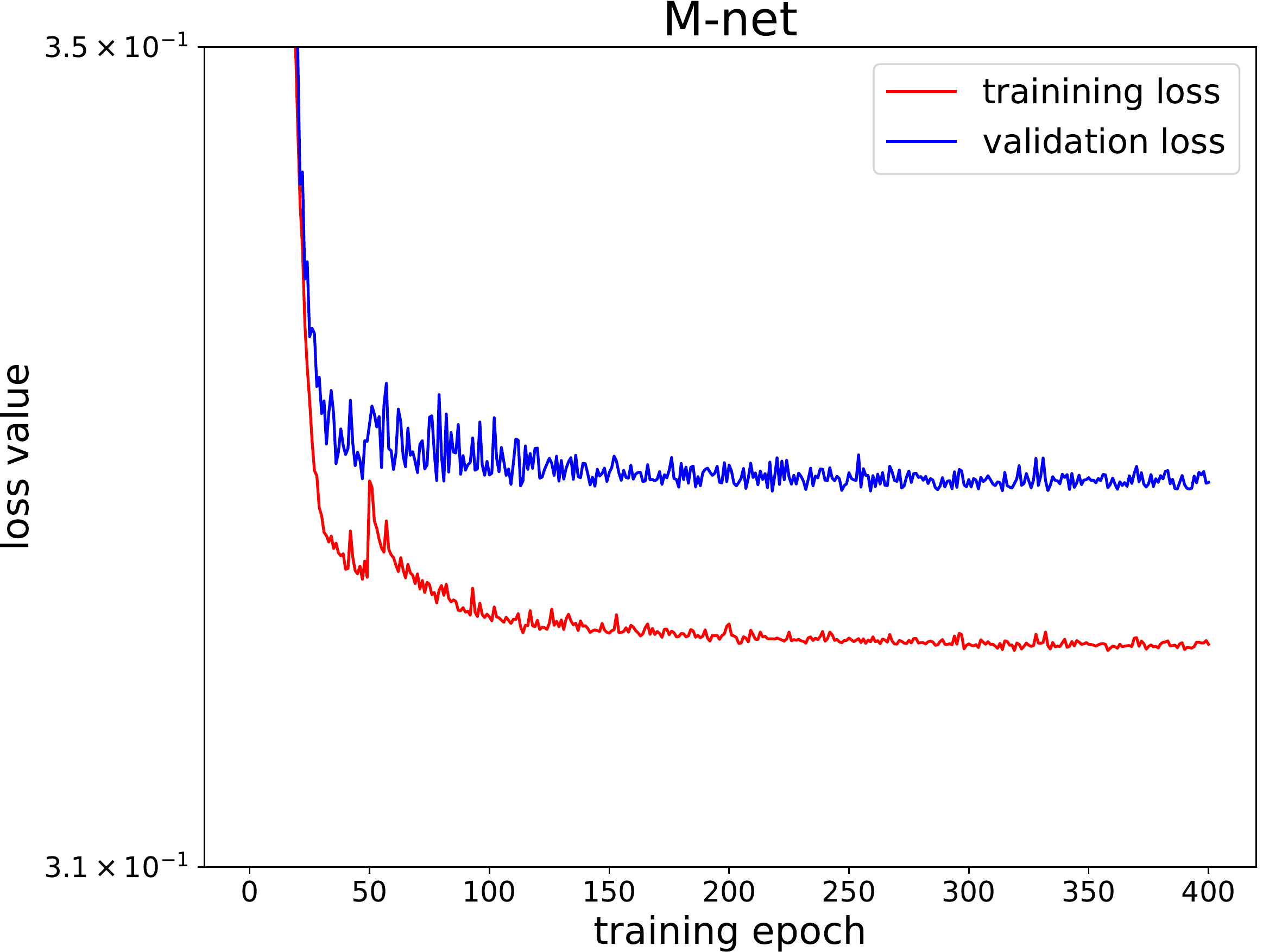}}
	\caption{Change of training loss and validation loss. (a) The loss of the proposed Pose-net. (b) The loss of the proposed M-net. }
	\label{fig:training_validation}
\end{figure}

We trained the Pose-net and the M-net for 700 and 400 epochs, respectively.
The stopping criterion was determined when validation loss stopped decreasing.
Fig. \ref{fig:training_validation} shows the training and validation loss for Pose-net and M-net.

\subsection{Results and Quantitative evaluations}

\subsubsection{Center position detection results}

For the quantitative evaluation of the Pose-net, we used the distance error between the output of the proposed method and ground-truth center positions in pixel space($3072\times1536$).
The error was computed for the case which succeed to detect the five lumbar vertebra correctly. The success rate was $96.25\%$ for 160 test data set.
Failure case which predicts the center position under five or wrong part(including T-spine and background) was excluded from the center position evaluation.
The distance error in the pixel space was shown in Table \ref{tbl:center_position_error}.
\begin{table}[!h]
	\footnotesize
	\centering
	\caption{Center position errors for the five lumber vertebra in pixel space are represented with mean and standard detivation.\label{tbl:center_position_error}}\vskip 0.1in
	\begin{tabular}{c|c|c|c|c|c}
		\multicolumn{6}{c}{\bf Distance Error in pixel space} \\ \hline
		{\bf L1} & {\bf L2} & {\bf L3} & {\bf L4} & {\bf L5} & {\bf All}\\ \hline
		 $26.84\pm10.58$ & $24.43\pm11.45$ & $24.57\pm10.63$ & $26.73\pm10.24$ & $24.17\pm11.18$ & $25.35\pm10.86$\\

	\end{tabular}
\end{table}

\begin{figure}[!h]
	\centering
	\subfigure[]{\includegraphics[height=0.32\textwidth]{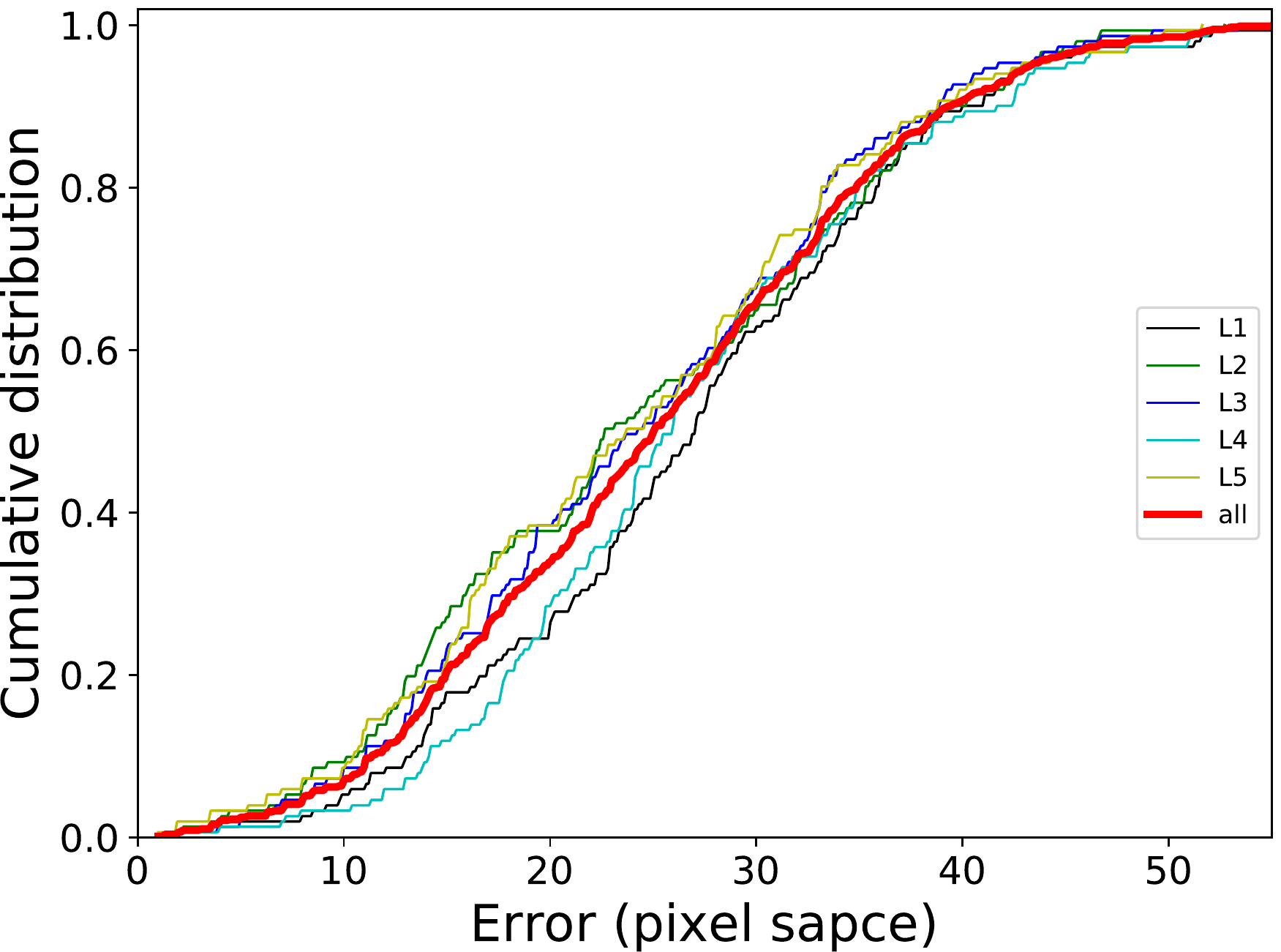}}
    \subfigure[]{\includegraphics[height=0.32\textwidth]{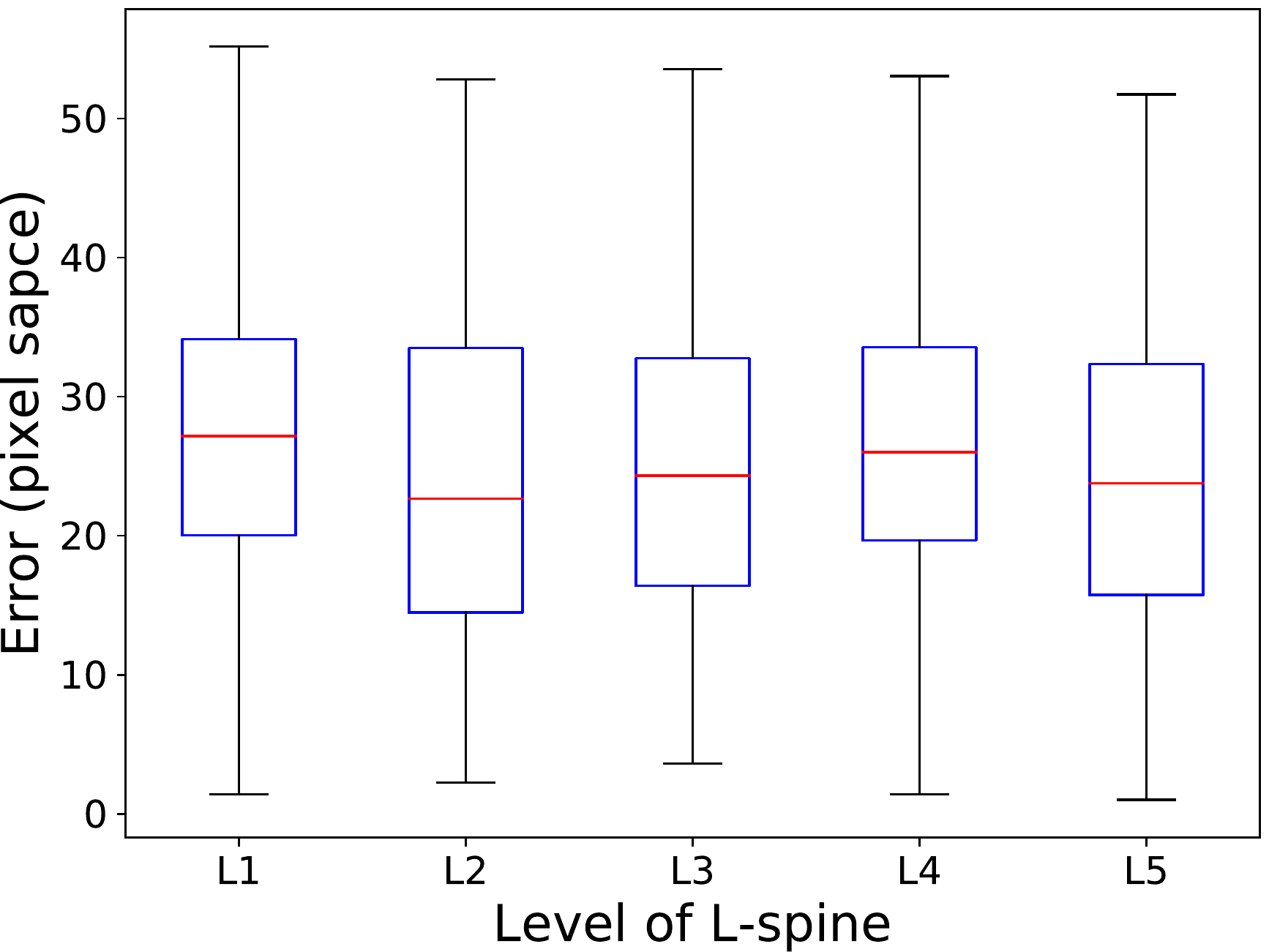}}
	\caption{Center position detection error of the proposed method. (a) Cumulative distribution of center position detection error. (b) Box plot of center position detection error.}
	\label{fig:center_position_error2}
\end{figure}

We then visualized the cumulative distribution of center position error for the five lumbar vertebra and all lumbar vertebra in Fig. \ref{fig:center_position_error2} (a).
From the cumulative distribution, it can be observed that most of center positions is within 50 pixels.
Fig. \ref{fig:center_position_error2} (b) shows the boxplot of the center position detection error.

For the qualitative evaluation of pose-estimation network, we visualized the confidence map and output of the center positions in Fig. \ref{fig:result_all} (b).

\subsubsection{Lumbar vertebra segmentation results}

\begin{figure}[h]	
	\centering
	{\includegraphics[width=0.9\textwidth]{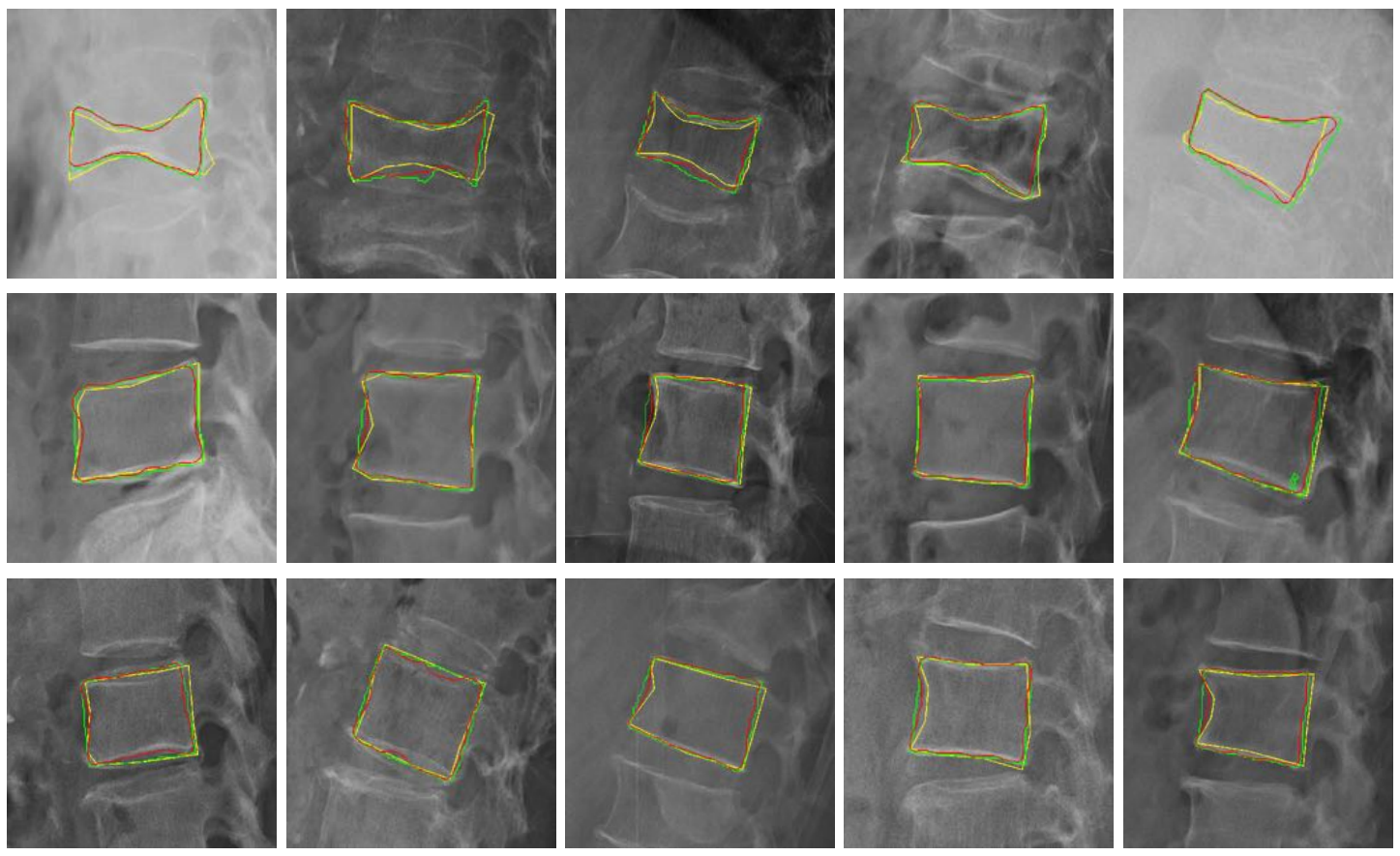}}
	\caption{The proposed segmentation results. Segmentation using M-net with Pose-net and fine tuning segmentation using level set were represented by green and red line, respectively. The yellow line denotes the ground-truth segmentation.}
	\label{fig:result_segmentation_lv_set}
\end{figure}

Fig. \ref{fig:result_segmentation_lv_set} shows the segmentation results using M-net with Pose-net and fine-tuning segmentation using level set.

For the evaluation of the proposed segmentation method, we used the region-based metric \cite{Udupa2006} including Dice similarity metric, precision, sensitivity, and specificity. The Dice similarity metric($d_{D}$) and the precision($d_{P}$) between $O_{\mbox{GT}}$ and $O_{\mbox{SG}}$ are defined as following:
\begin{equation}
d_{D}:= \frac{2|O_{\mbox{GT}}\cap O_{\mbox{SG}}|}{|O_{\mbox{GT}}|+|O_{\mbox{SG}}|}, d_{P}:=\frac{|O_{\mbox{GT}}\cap O_{\mbox{SG}}|}{|O_{\mbox{GT}}\cup O_{\mbox{SG}}|}
\end{equation}
where $O_{\mbox{SG}}$ is the lumbar vertebra region obtained from segmentation and $O_{\mbox{GT}}$ is the ground-truth segmentation. Here, $d_D$ describes how the ground-truth and detected region are close to and overlapped with each other.
The the sensitivity($d_{Sen}$) and the specificity($d_{Spe}$) are defined as
\begin{equation}
	d_{Sen}:=\frac{|O_{\mbox{GT}}\cap O_{\mbox{SG}}|}{|O_{\mbox{GT}}|}, d_{Spe}:=\frac{|(O_{\mbox{GT}}\cup O_{\mbox{SG}})^c|}{|(O_{\mbox{GT}})^c|}.
\end{equation}

We compared the multi-step proposed method with existing deep learning segmentation method.
For comparison, we used M-net which take an image $I\in{\mathbb R}^{512\times256}$ as an input and produce an output of size $512\times256$ for segmentation of the five lumbar vertebra.
We will refer this M-net as original M-net to distinguish the M-net used in the proposed method.
We should note that the M-net used in the proposed method takes as input the extracted patch $I_p$ to segment the individual lumbar vertebra.
We also used U-net\cite{Ronneberger2015} for both the proposed method and existing method, namely original U-net.
The evaluation results are reported in Table \ref{tbl:comparison_segmentation}. Here, Pose-net+M-net+Level set denotes the fine-tuning segmentation and Pose-net+M-net represents the segmentation without fine-tuning.

\begin{table}[!h]
	\footnotesize
	\centering
	\caption{Comparison the result for several methods. Evaluation of segmentation results using multiple metrics. Dice coefficient, precision, sensitivity, and specificity are represented with mean and standard detivation.\label{tbl:comparison_segmentation}}\vskip 0.1in
	\begin{tabular}{c||cccc}
		&\multicolumn{4}{c}{\bf Region-Based Metric($\%$)} \\ \hline
		\bf Method & \bf Dice & \bf Precision & \bf Sensitivity & \bf Specificity \\ \hline
		\bf Pose-net+M-net+Level set& $91.60\pm2.22$ & $84.57\pm3.64$ & $90.13\pm2.91$ & $99.59\pm0.17$  \\ \hline
		\bf Poes-net+U-net+Level set & $91.05\pm3.50$ & $83.74\pm5.47$ & $90.76\pm3.72$ & $99.49\pm0.26$ \\ \hline
		\bf Pose-net+M-net& $90.38\pm4.31$ & $82.72\pm6.86$ & $92.74\pm4.33$ & $99.32\pm0.33$ \\ \hline
		\bf Poes-net+U-net & $90.14\pm4.26$ & $82.32\pm6.79$ & $93.61\pm3.69$ & $99.24\pm0.035$\\ \hline
		\bf Original M-net & $88.31\pm5.97$ & $79.55\pm9.01$ & $89.22\pm7.27$ & $99.31\pm0.49$ \\ \hline
		\bf Original U-net & $87.39\pm7.13$ & $78.27\pm10.46$ & $89.07\pm9.01$ & $99.22\pm0.46$ \\
	\end{tabular}
\end{table}

From this results we can see that the proposed method achieves the improved Dice similarity metric by combining deep learning method and level-set method.
The level-set method combined with segmentation deep learning takes advantages of clear edge at anterior wall, upper, and lower plate of vertebra body, therefore Dice similarity metric was increased by reducing false-positive of segmentation.
However, the posterior wall of vertebral body has unclear boundary due to overlapping two pedicles in the lateral view of lumbar X-ray image, it is difficult to capture the boundary of the posterior wall using level set.
This causes level set method to segment inside region of vertebral body, resulting in decreasing of sensitivity value.
We expected that one can improve the level set method to achieve the more accurate segmentation, but this is out of the scope of the our paper.

\begin{figure}[h]	
	\centering
	{\includegraphics[width=0.9\textwidth]{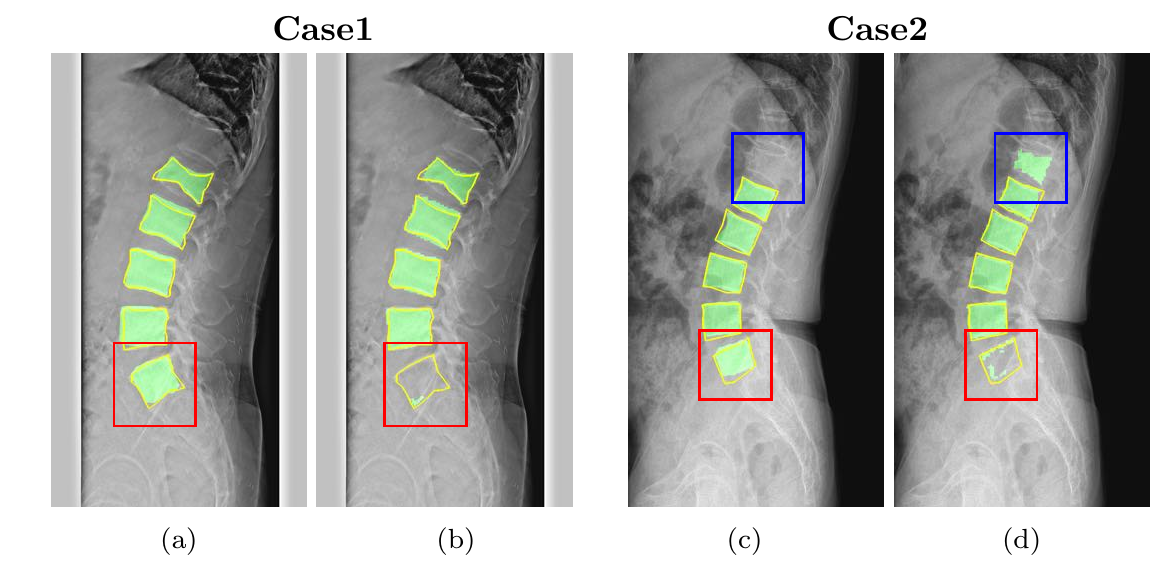}}
    \caption{Comparison of the proposed method and the existing method. (a) and (c) show the segmentation using M-net with Pose-net. (b) and (d) show the inaccurate segmentation using original M-net. The yellow line denotes the ground-truth. The red box describes the L5 vertebra and the blue box describes the T12 vertebra.}
    \label{fig:result_segmentation_comp}
\end{figure}

Fig. \ref{fig:result_segmentation_comp} shows the comparison results with two cases. From red box in Fig. \ref{fig:result_segmentation_comp}, we can observed that the original M-net failed to segment L5 vertebra while the proposed method can segment L5 vertebra by taking advantage of accurate localization of the spine.
This accurate localization can also prevent segmenting of thoracic spine. See blue box in Fig. \ref{fig:result_segmentation_comp} (c) and (d).

\begin{figure}[h]	
	\centering
	{\includegraphics[width=0.9\textwidth]{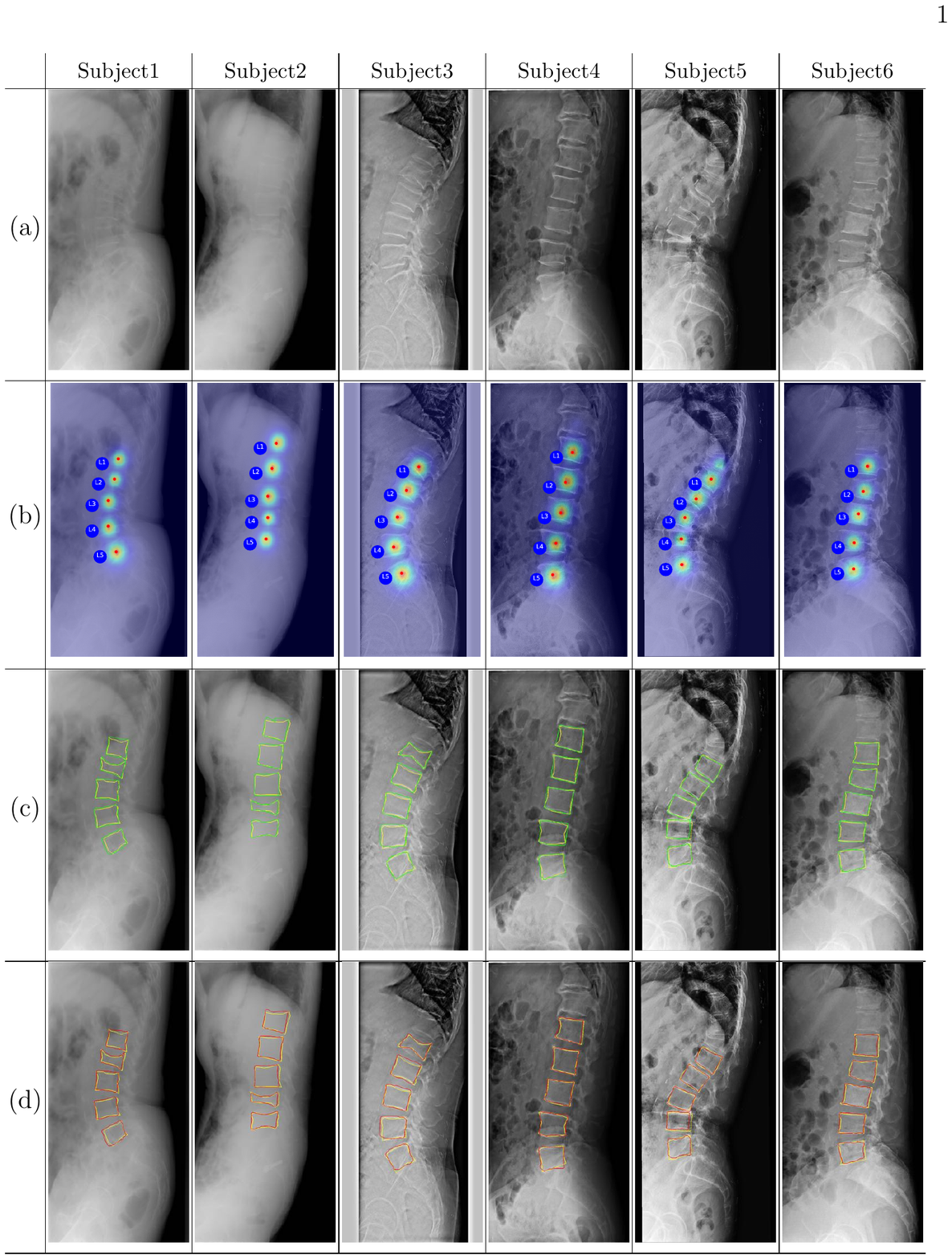}}
	\caption{The results of the proposed method for six subject. (a) Original X-ray images. (b) Confidence map and detected center positions of the lumbar vertebra. (c) Segmentation result from M-net. (d) Fine tuning segmentation using level set.}
	\label{fig:result_all}
\end{figure}

The results of the entire process for selected six subject were shown in Fig. \ref{fig:result_all}.

\begin{figure}[h]
	\centering
\subfigure[]{\includegraphics[height=0.45\textwidth]{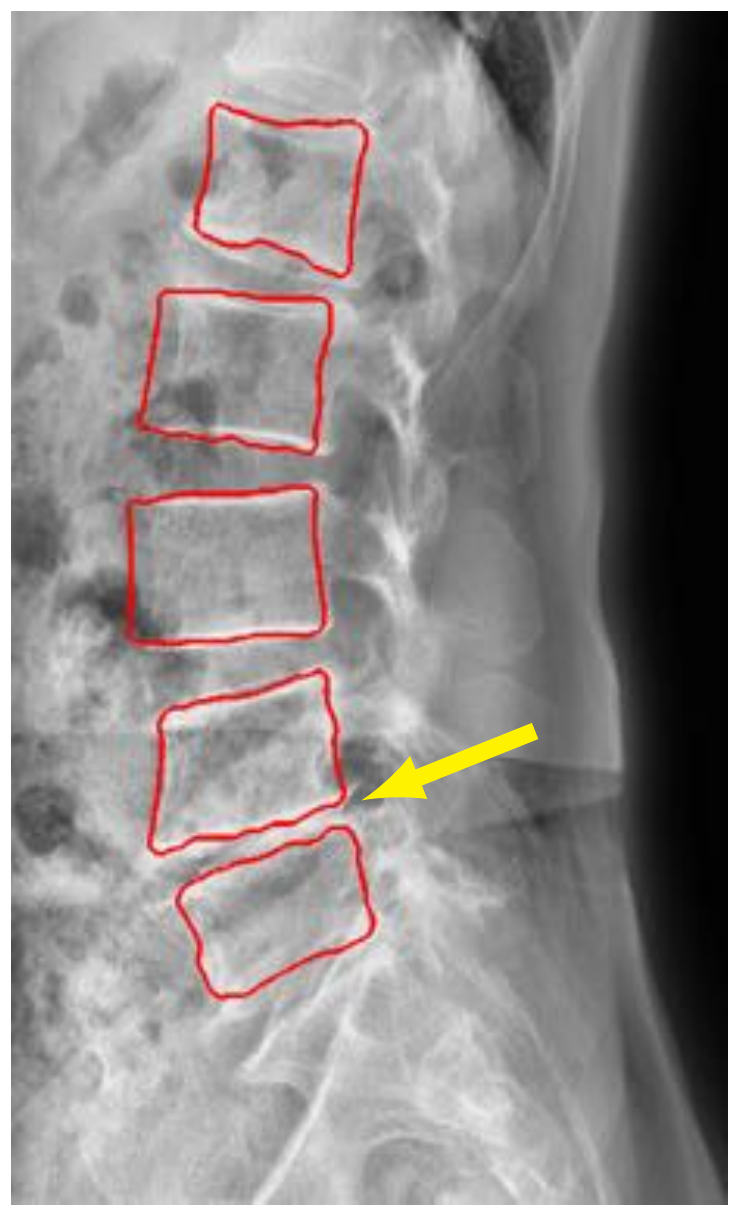}}
\subfigure[]{\includegraphics[height=0.45\textwidth]{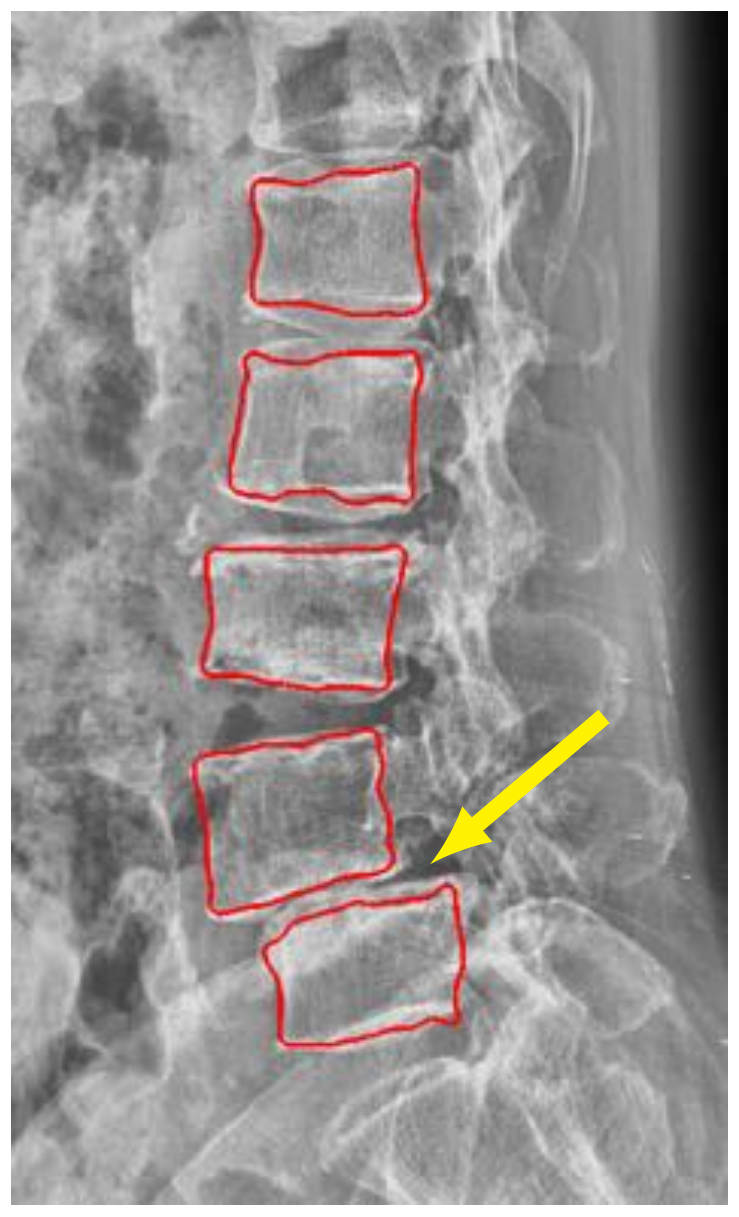}}
	\caption{Disc space narrowing and spondylolisthesis in lateral view of lumbar spine. Decreased L4-5 disc space height (a) and anterior displacement of the L4 vertebra over the L5 (b) were denoted by yellow arrow. Here, red contour denotes the segmentation results. }
	\label{fig:spondylolisthesis}
\end{figure}

\section{Discussion and Conclusion}
The main contribution of the proposed method is that it achieves (i) accurate and robust identification of each lumbar vertebra using a pose-driven deep-learning technique, and (ii) fine segmentation of individual vertebra using a hierarchical method that combines M-net and level-set methods.

Lumber compression fractures are becoming increasingly prevalent in Korea as the incidence of osteoporosis increase with aging populations. Compression fracture is the most common fracture in osteoporosis patients. In Korea, the burden of medical imaging due to aging is increasing rapidly, and the rate of increase of radiologists is falling short of that. As a result, radiologists are more likely to be difficult to read quickly and accurately. In particular, if imaging diagnosis is missed or delayed in spinal compression fractures, it can lead to complications such as height reduction and scoliosis. Therefore, the automatic vertebral segmentation could play an important role in improving physicians' workflow with being diagnosed quickly and accurately through images.

The automatic vertebral segmentation may proceed with follow-up studies in the automatic grading of compression fracture in place of existing semiquantitative grade system(genant grade). Such an automatic quantitative grading method would result in a clear and reproducible definition of compression fracture.
In addition to compression fracture in lumbar vertebra, automatic vertebral  segmentation study can also enable research on other diseases such as degenerative changes(including disc space narrowing and degenerative spondylolisthesis as shown in Fig. \ref{fig:spondylolisthesis}) and traumatic conditions such as including burst fracture. It is also believed that studies of various diseases in the spine (bone tumor such as metastasis, infectious disease such as pyogenic spondylitis, autoimmune disease such as Ankylosing spondylitis, etc.) could be possible if they were extended to cervicothoracic spine, sacrum, and coccyx.

\section*{Acknowledgement}
This work was supported by Samsung Science $\&$ Technology Foundation (No. SSTF-BA1402-01).
K.C.K. was supported by NRF grant 2017R1E1A1A03070653.

\section*{References}

\bibliographystyle{dcu}

\end{document}